\documentstyle[epsfig,prd,aps,twocolumn,amsmath]{revtex}

\newcommand{\psibar}{{\psi}^\dag}
\newcommand{\psiLe}{{\psi_{\mathrm{L}}^{\mathrm{e}}}}
\newcommand{\psiLo}{{\psi_{\mathrm{L}}^{\mathrm{o}}}}
\newcommand{\psiRe}{{\psi_{\mathrm{R}}^{\mathrm{e}}}}
\newcommand{\psiRo}{{\psi_{\mathrm{R}}^{\mathrm{o}}}}
\newcommand{\psiRLeo}{{\psi_{\mathrm{R},\mathrm{L}}
    ^{\mathrm{e},\mathrm{o}}}}

\newcommand{\e}{{\mathrm{e}}}
\newcommand{\dotmu}{{\dot{\mu}}}
\newcommand{\halb}{\frac{1}{2}}
\newcommand{\Tr}{{\mathrm{Tr}}}
\newcommand{\Z}{{\mathcal{Z}}}
\newcommand{\D}{{\mathcal{D}}}
\renewcommand{\i}{{\mathrm{i}}}
\newcommand{\Asl}{{\rlap{\,/}{{A}}}}
\newcommand{\Dsl}{{\rlap{\,/}{{D}}}}

\renewcommand{\dag}{\dagger}

\begin{document}            % the beginning
\preprint{HEP-ph/9806372}
\draft

\title{Random Matrix Model for\\Wilson Fermions on the Lattice}
\author{Holger Hehl and Andreas Sch\"afer}
\address{Institut f\"ur Theoretische Physik, Universit\"at Regensburg,\\
         Universit\"atsstr. 31, D-93040 Regensburg, Germany}
\date{February 3, 1999}
\maketitle
\begin{abstract}
  We describe a random matrix model suitable for the simulation of the
  eigenvalues of the Dirac operator on the lattice for Wilson
  fermions. We compare the obtained global eigenvalue spectrum for
  various values of the hopping parameter $\kappa$ with lattice
  results of Kalkreuter. The agreement is surprisingly good.
\end{abstract}

\narrowtext\flushbottom

Recently, it became clear that the microscopic spectral properties of
the lattice QCD Dirac operator are universal and can be reproduced by
simple models that only share basic symmetry properties with real QCD
\cite{Ver97,Leu92}. Such models are provided by random matrix theory
(RMT) \cite{Ver97,Guh98}. This universality has recently been
demonstrated for the staggered lattice Dirac operator in quenched
\cite{Ber97a} and unquenched \cite{unquenched} SU(2). Since the
Banks--Casher formula $\Sigma = \pi\rho(0)/V$ \cite{Ban80} links the
spectral density at zero virtuality to the chiral condensate $\Sigma$
the distribution of the small eigenvalues is of great importance for,
e.g., the understanding of the chiral phase transition. RMT has also
solved a long standing problem of lattice calculations at finite
chemical potential \cite{Ste97}. Recently, also the predictions for
the energy scale at which RMT and lattice QCD start to deviate was
confirmed, which among others tested the validity of the
Gell-Mann--Oakes--Renner relation on the lattice \cite{Osb98}. While
RMT makes reliable predictions only for microscopic spectral
fluctuations, these successes encourage us to see if the global
spectral properties can be reproduced. This is the aim of our
contribution for the case of Wilson fermions. Since the universality
argument only applies to microscopic properties one cannot hope to fit
global spectra except for very special cases. We analyze such a case,
namely, gauge theories with infinitely strong coupling, i.e.,
$\beta\to 0$.  Such systems can be hoped to be sufficiently chaotic as
to show random matrix characteristics even on global scales.
Furthermore our model studies suggest that to describe Wilson fermions
a $4\times 4$ block structure is needed in random matrix theory. We
expect this to hold true also for a description of microscopic
properties at non-zero $\beta$. Random matrix models for Wilson
fermions at non-zero $\beta$ do not yet exist, but are of great
practical importance.

{\samepage We start our analysis with the Euclidean action of lattice
SU(2) theory with Wilson fermions in the fundamental representation
which can be written as\pagebreak
\begin{equation} \label{SE}
\begin{split}
S_{E}&=\frac{1}{2\kappa}\sum_n\psibar(n)\psi(n) \\
&\quad\mbox{}-\halb\sum_{n,\mu}\bigl[\psibar(n)(r-\gamma_\mu) 
U_\mu(n)\psi(n+\mu) \\
&\qquad\qquad\quad\mbox{}+\psibar(n+\mu)(r+\gamma_\mu)
U_\mu^\dag(n)\psi(n)\bigr]
\\
&\quad\mbox{}+\frac{4}{g^2}\sum_P\Bigl\{1-\frac{1}{4}
\Tr\bigl[U_P(n)+U_P^\dag(n)\bigr]\Bigr\}
\end{split}
\end{equation}
with the Wilson parameter $r$ that we set to $1$ in the following
\cite{Rothe}. The gauge fields $A_\mu$ are contained in the link
variables $U$. With this action a gauge invariant partition function
can be constructed from which one can obtain vacuum expectation values
of operator products in the usual way. In these partition functions we
average over all gauge field configurations.}

In random matrix theory we substitute the Dirac operator which
includes the gauge fields by random matrices of a particular ensemble
to model the very strong fluctuations of the Dirac operator when
calculated with lattice gauge theory. The integration is then
performed over the independent entries of the matrices \cite{Ver97}.
With this approximation the gluons decouple from the quarks and can be
integrated out, i.e., they can be neglected in random matrix models if
one is only interested in quark observables. The symmetry properties
of the random matrix depend on the underlying gauge group and the
fermion representation. Usually in QCD we are dealing with fermions in
the fundamental representation of the SU(3) gauge group in which case
one has to use the Gaussian Unitary Ensemble (GUE). In order to
compare our results with Kalkreuter \cite{Kal96} who investigated the
operator $\gamma_5(\Dsl +m)$ for massive Wilson fermions in an SU(2)
gauge field background we have to use matrices of the Gaussian
Orthogonal Ensemble (GOE) \cite{Ver94}. The Euclidean partition
function with two random matrices can be written according to the
above arguments as
\begin{equation} \label{Z} \Z = \int\D [A,B] \e^{-\Sigma_A A^\dag
  A -\Sigma_B B^\dag B}\int\D [\psibar,\psi]
\e^{-\psi^\dag({\not D} +m)\psi}
\end{equation}
with the parameters $\Sigma_A$ and $\Sigma_B$ that scale the
distribution variance of the Gaussian Ensembles. We will now specify
the operator $\Dsl +m$ in Eq.~\eqref{Dsl} and explain why we are
using two different random matrices.

In the following we separate the Dirac spinors into left- and
right-handed fields $\psi_{\mathrm{L,R}} = \halb(1\mp\gamma_5)\psi$
with the Euclidean $\gamma_5 = -\gamma_5^{\mathrm{M}}$. Furthermore we
distinguish between \emph{even} and \emph{odd} lattice sites. These
two sub-lattices couple rather independently (the two groups are often
called `red' and `black' lattice sites in analogy to the colors of a
checkerboard). The spinor field is then written as
$(\psiRe,\psiLe,\psiRo,\psiLo)$, where $\psiRLeo =~
\bigl(\psiRLeo(1),\dots,\psiRLeo(\frac{N}{2})\bigr)$ are vectors with
respect to the lattice sites $x_\mu(1),\dots,x_\mu(N)$.

In the above basis the mass term of \eqref{SE} is simply a $4\times 4$
block diagonal matrix and the interaction term becomes
\begin{subequations}
\begin{align}
&-\halb(\psiLe^\dag\psiLo+\psiRe^\dag\psiRo + \psiLo^\dag\psiLe +
\psiRo^\dag\psiRe) \label{a}\\
&-\halb(-\psiLe^\dag\gamma_\dotmu\psiRo
-\psiRe^\dag\gamma_\dotmu\psiLo +\psiLo^\dag\gamma_\dotmu\psiRe
+\psiRo^\dag\gamma_\dotmu\psiLe) \label{b}\\
&-\frac{\i}{2}ga(\psiLe^\dag A_\dotmu\psiLo +\psiRe^\dag
A_\dotmu\psiRo -\psiLo^\dag A_\dotmu^\dag\psiLe -\psiRo^\dag
A_\dotmu^\dag\psiRe) \label{c}\\
&-\frac{\i}{2}ga(-\psiLe^\dag\Asl\psiRo -\psiRe^\dag\Asl\psiLo
-\psiLo^\dag\Asl^\dag\psiRe -\psiRo^\dag\Asl^\dag\psiLe). \label{d}
\end{align}
\end{subequations}
The notation $\dotmu$ indicates that $\mu$ is not a free Lorentz index
but is contracted with a corresponding one that is hidden in the
$\psi$'s. We have expanded the link variables $U_\dotmu$ to first
order in $A_\dotmu$, i.e., $U_\dotmu = 1+\i gaA_\dotmu$. The term
\eqref{a} then leads to the constant block matrix
\begin{subequations} \label{parts}
\begin{equation}
\begin{pmatrix}0&0&1&0\\0&0&0&1\\
    1&0&0&0\\0&1&0&0\end{pmatrix}.
\end{equation}
In \eqref{c} we replace $A_\dotmu$ with $\i/ga$ times the random
matrix $A$ such that we get the block matrix
\begin{equation}
\begin{pmatrix}0&0&-A&0\\0&0&0&-A\\
    -A^\dag&0&0&0\\0&-A^\dag&0&0\end{pmatrix}.
\end{equation}
In \eqref{d} we do the same with $\Asl$ but calling the random matrix
$B$ which gives us
\begin{equation}
\begin{pmatrix}0&0&0&+B\\0&0&+B&0\\
    0&-B^\dag&0&0\\-B^\dag&0&0&0\end{pmatrix}.
\end{equation}
\end{subequations}
In the remaining term \eqref{b} which has the same structure as
\eqref{d} we are faced with a current-like form, bilinear in the
fields. As in \eqref{c} different $\gamma$ matrices are used depending
on the lattice sites which makes this term sufficiently random to be
absorbed in the matrix $B$. To obtain the above matrices we have
rescaled the Dirac fields by $1/\sqrt{-1/2}$.

The expansion of $U_\dotmu$ we used to argue for our assumed block
structure is justified in the perturbative domain of small $g$. Here
we are rather interested in the ergodic regime of fully developed
chaoticity, which corresponds to rather large $g$. However, we are
only interested in the general symmetry properties, which are the same
in both regimes.

Adding up the above contributions \eqref{parts} we arrive at the
following expression for the complete Dirac operator:
\begin{equation} \label{Dsl}
\Dsl +m= \begin{pmatrix}1/2{\kappa}&0&1-A&B\\
    0&1/{2\kappa}&B&1-A\\
  1-A^\dag&-B^\dag&1/{2\kappa}&0\\
  -B^\dag&1-A^\dag&0&1/{2\kappa}\end{pmatrix}.
\end{equation}

Neglecting the distinction between even and odd fields and looking at
the chiral structure only, $\Dsl$ has the structure of a $2\times 2$
block matrix, namely,
$\left(\begin{smallmatrix}0&*\\\ast&0\end{smallmatrix}\right)$ since
it is anti-commuting with $\gamma_5$. This ensures that the
eigenvalues of $\Dsl$ are distributed around zero symmetrically. Let
us stress that we have in no way derived Eq.~\eqref{Dsl}. We basically
only give some hand-waving arguments for the form in Eq.~\eqref{g5Dsl}
which is then tested against the numerical data.  Definitely in the
end more realistic models will differ from \eqref{Dsl} but we believe
that they will share the $4\times 4$ structure.

Kalkreuter investigated in his analyses the operator
$\gamma_5(\Dsl+m)$ so we have to multiply \eqref{Dsl} by $\gamma_5$.
Here we must carefully keep in mind that the $B$ matrix resulted from
the $\Asl$ term which implicitly contains the $\gamma$ matrices. When
we also define our spinor basis according to
$(\psiRe,-\psiLe,\psiRo,\psiLo)$ we finally arrive at
\begin{equation}
\label{g5Dsl} \gamma_5(\Dsl +m)=
\begin{pmatrix}1/{2\kappa}&0&1-A&B\\
  0&-1/{2\kappa}&B&1-A\\1-A^\dag&B^\dag&1/{2\kappa}&0\\
  B^\dag&1-A^\dag&0&-1/{2\kappa}\end{pmatrix}.
\end{equation}

At first we tried to determine the spectral density of the above
operator analytically. For this purpose we chose \eqref{g5Dsl} as the
matrix in the determinant of \eqref{Z} and integrated out the
$2\cdot\halb\frac{N}{2}(\frac{N}{2}+1)$ real matrix variables by means
of the Hubbard--Stratonovich transformation which lead to a partition
function with integration over only $8$ complex variables. It turned
out that the resulting saddle point equations are too complicated to
be solved analytically so that we decided to diagonalize the operator
matrix numerically.

\begin{figure}[htb]
  \centering
  \begin{minipage}[b]{0.4\textwidth}
    \centering
     \epsfig{file=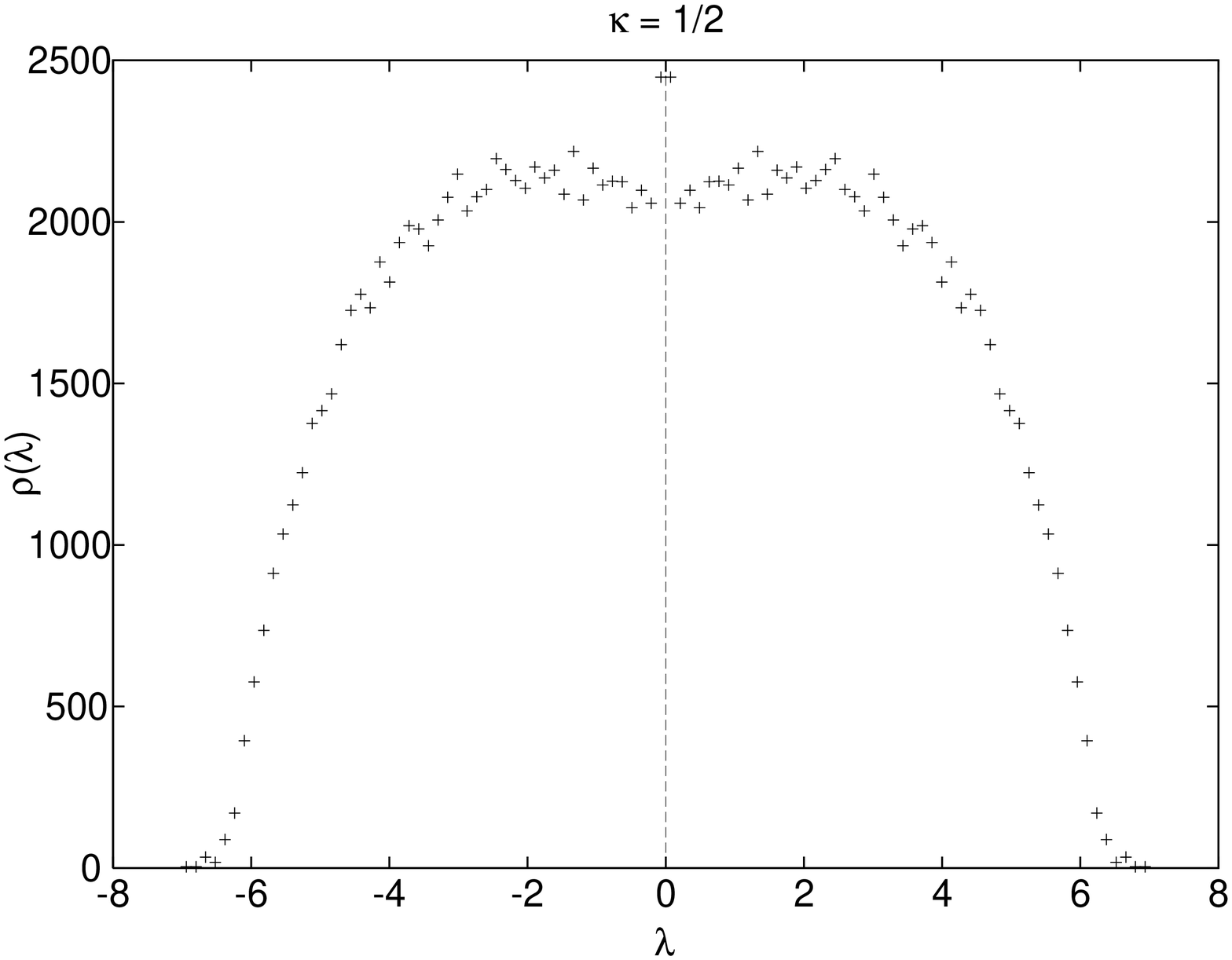, height=5.4cm}
     \epsfig{file=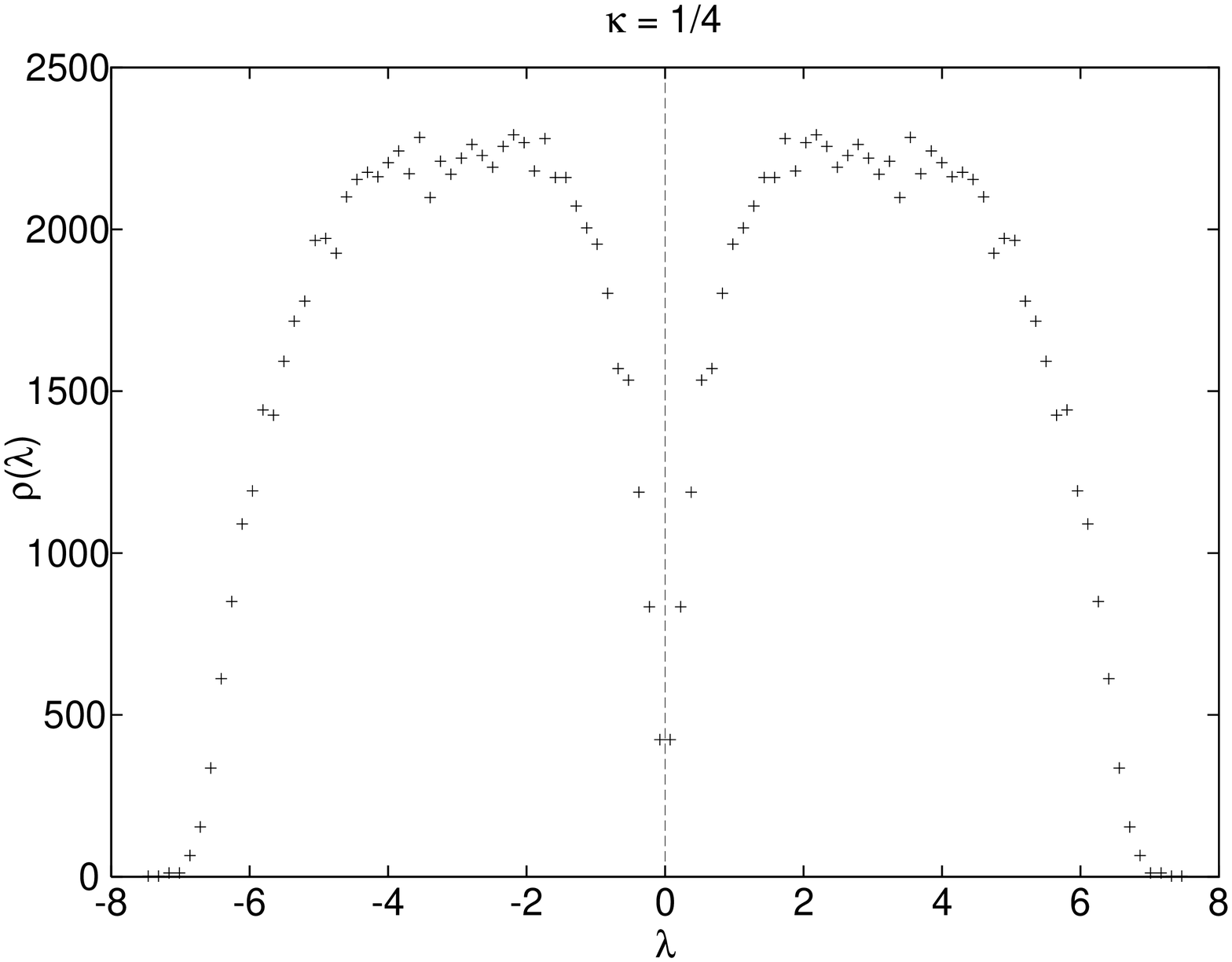, height=5.4cm}
     \epsfig{file=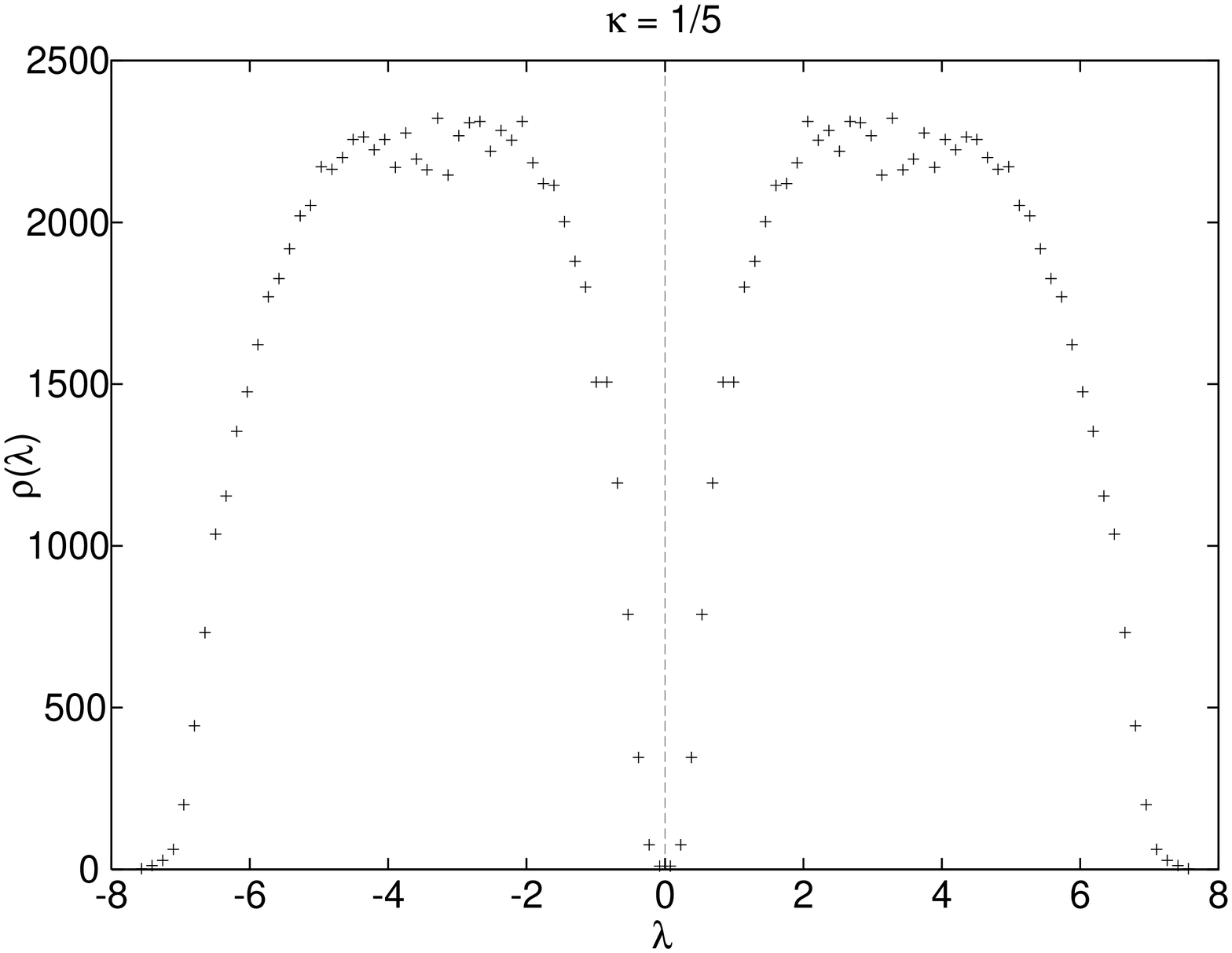, height=5.4cm}
     \epsfig{file=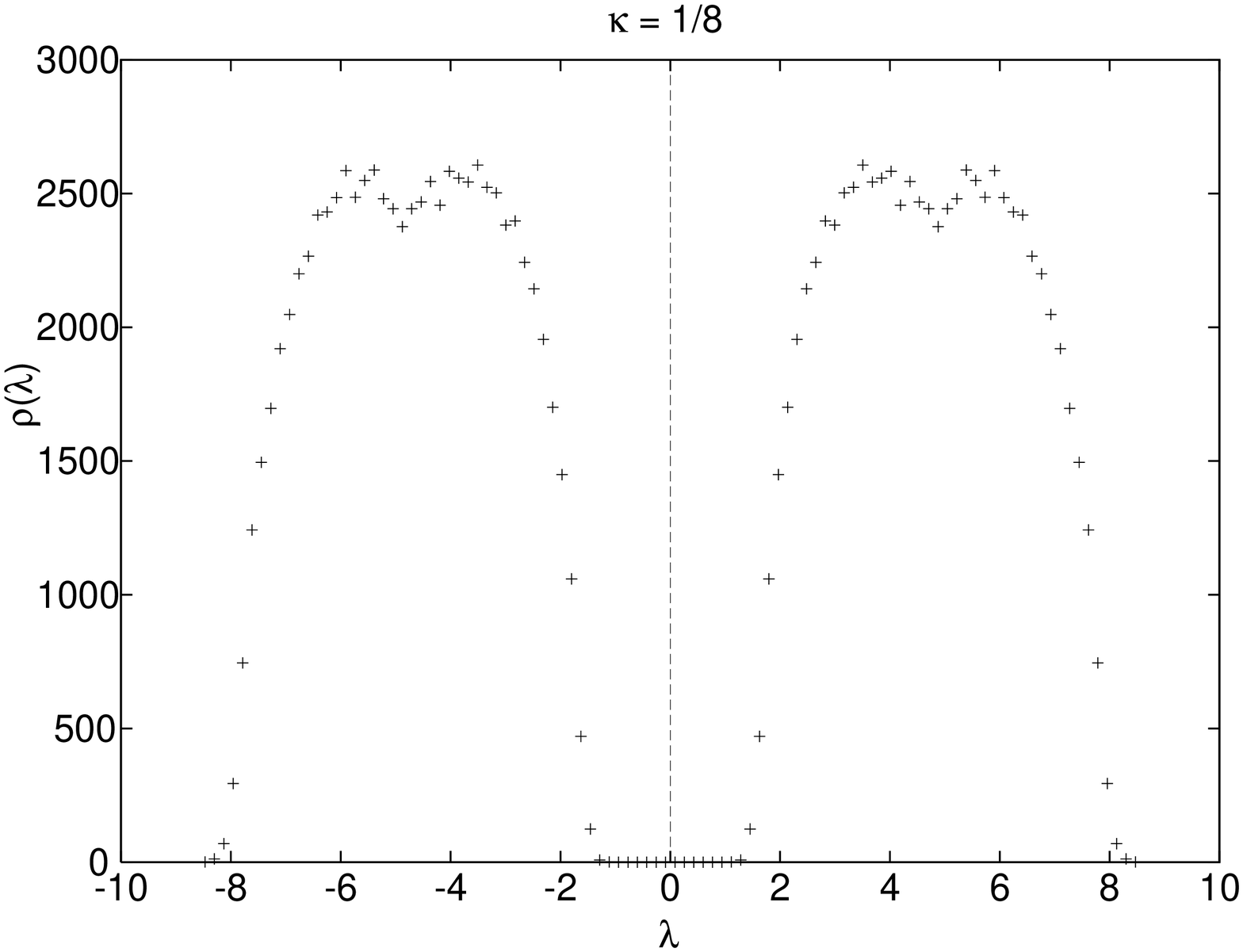, height=5.4cm}
  \end{minipage}\hspace{0.05\textwidth}
  \begin{minipage}[b]{0.4\textwidth}
    \centering
     \epsfig{file=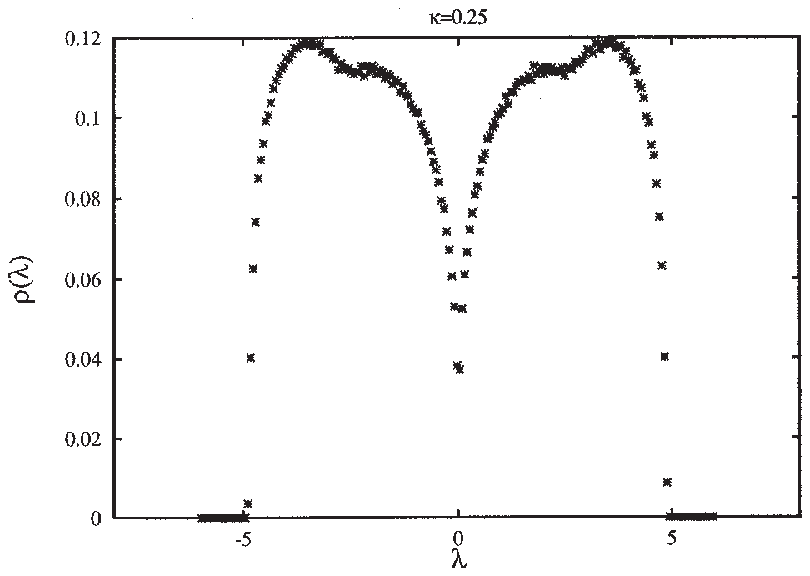, height=5.4cm, width=7cm}
     \epsfig{file=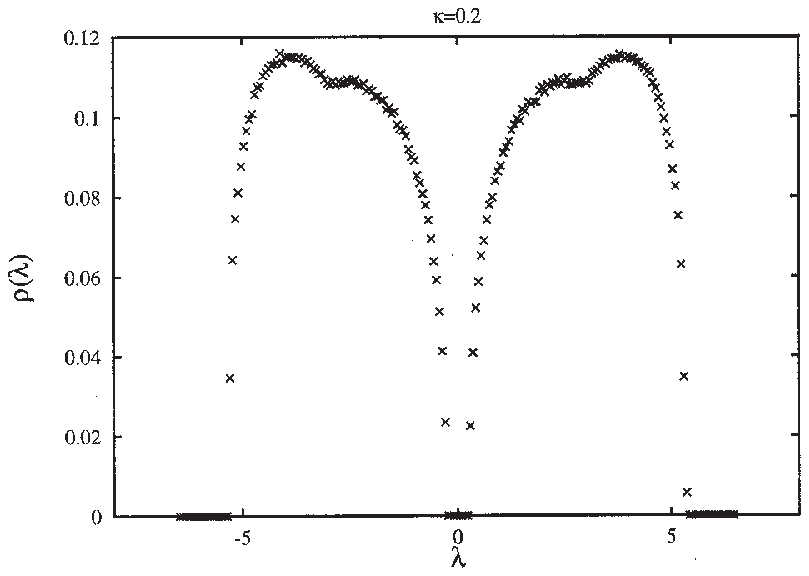, height=5.4cm, width=7cm}
     \epsfig{file=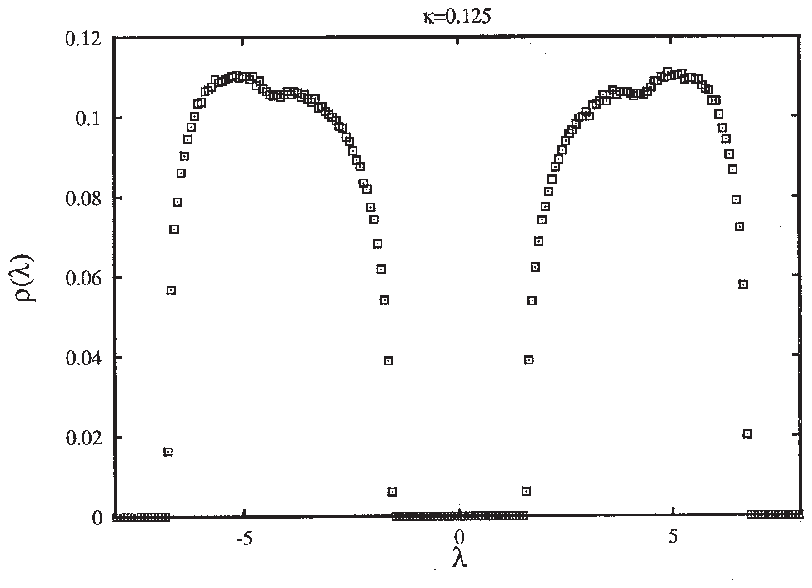, height=5.4cm, width=7cm}
  \end{minipage}
 \caption{spectral densities of the massive Dirac operator for
   various values of the hopping parameter $\kappa$. On the right hand
   side are shown Kalkreuter's results for $\beta=0.0$ extracted from
   the first picture of Fig.~\protect\ref{beta18}, but with eigenvalue
   normalization as on the left hand side. The chosen parameters on
   the left are $\Sigma_A=2/25$, $\Sigma_B=8/25$. The hopping
   parameter is $\kappa=1/2$, $1/4$, $1/5$, and $1/8$, respectively.}
  \label{fig}
\end{figure}

We calculated the eigenvalues of this matrix for a large number of
random variables matching the selected ensemble (GOE). We display the
results in histograms for several values of the parameter $\kappa =
(2m+8)^{-1}$ (see Fig.~\ref{fig}). Our results are on the left side of
this figure. They are compared with the corresponding $\beta=0$ data
of Kalkreuter taken from \cite{Jur96} which are shown on the right
side.\pagebreak

\widetext

\clearpage
\narrowtext
\begin{figure}[t]
  \centering
    \epsfig{file=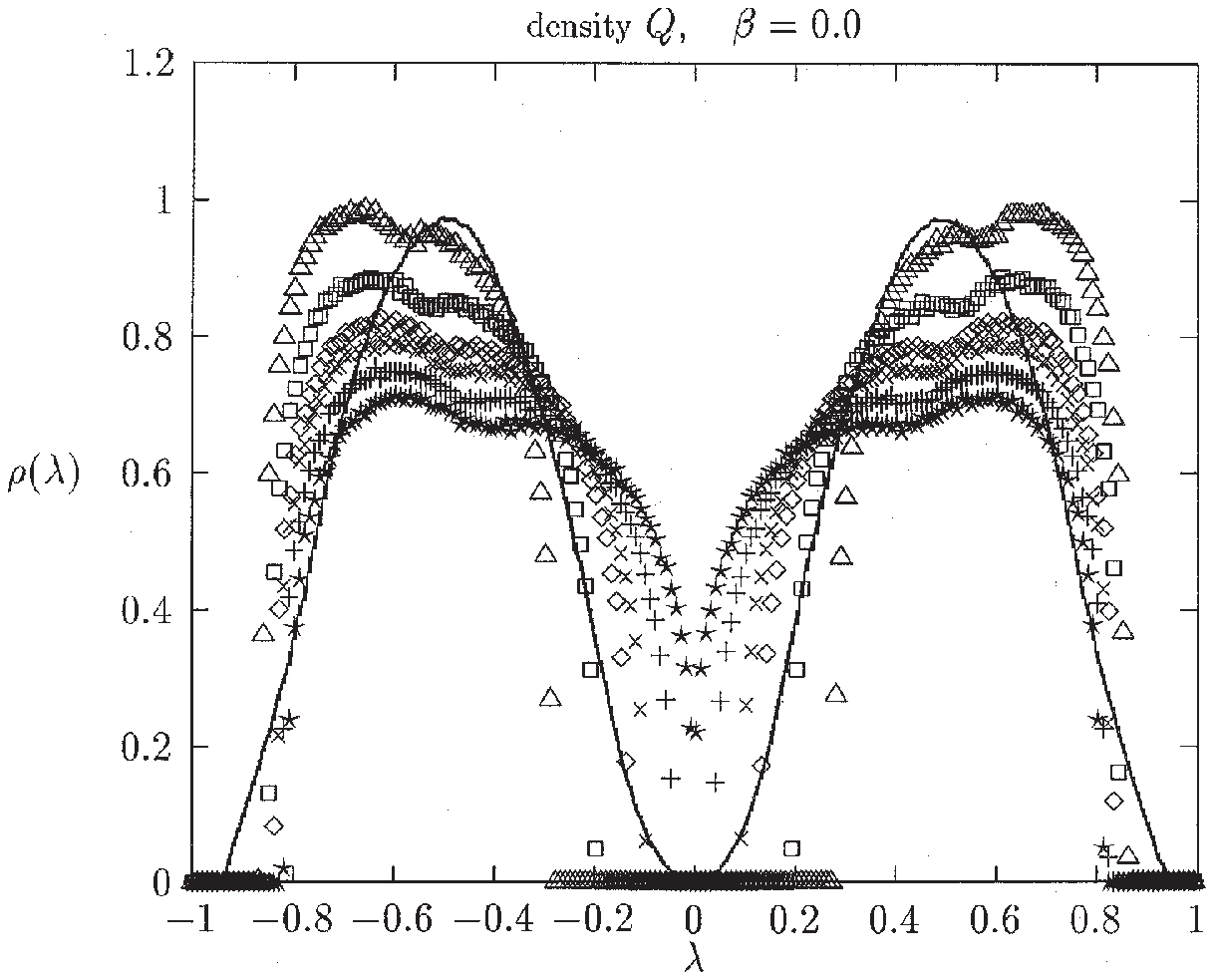,height=7.0cm} %,width=0.47\textwidth}
    \epsfig{file=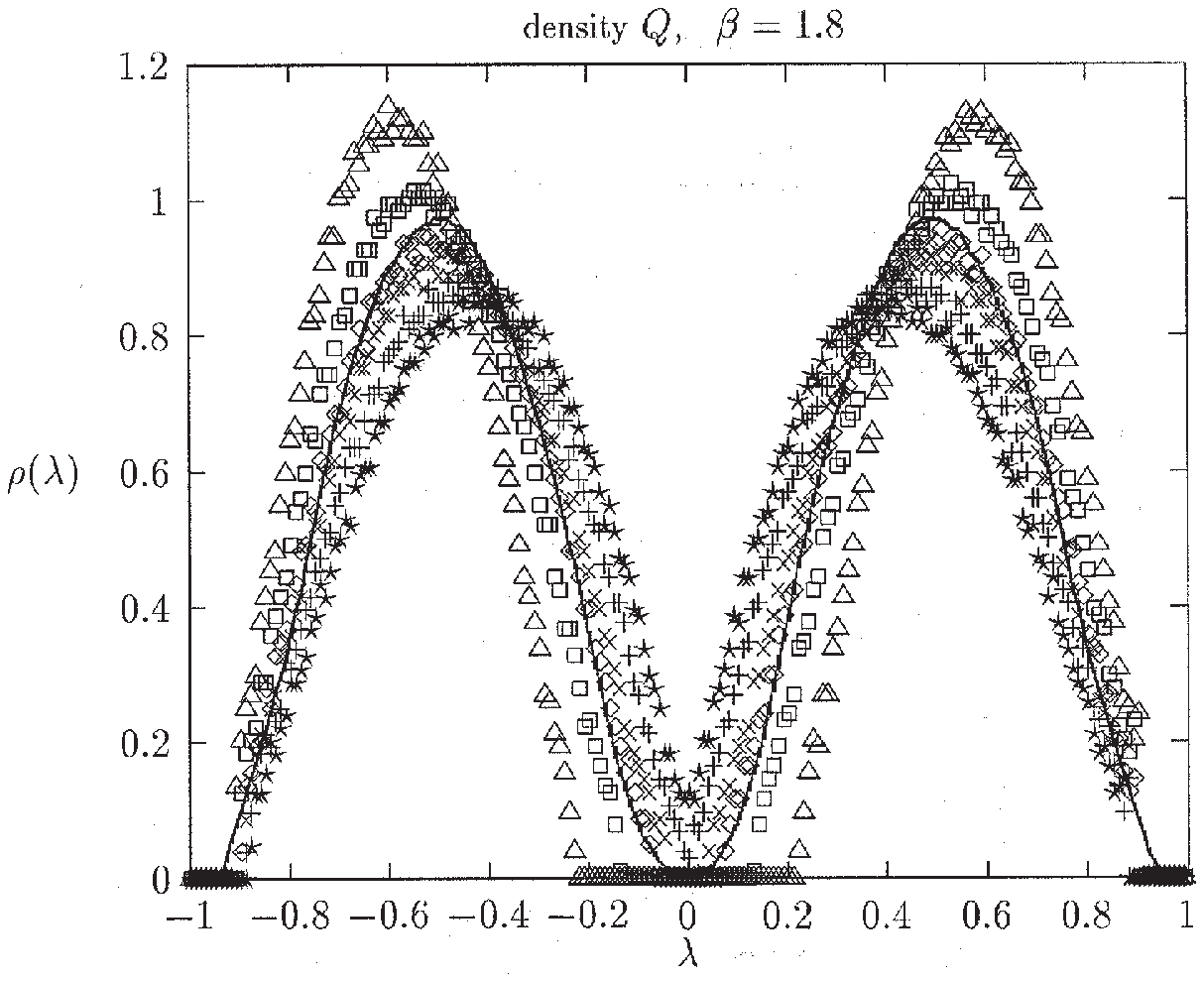,height=7.0cm} %,width=0.47\textwidth}
    \epsfig{file=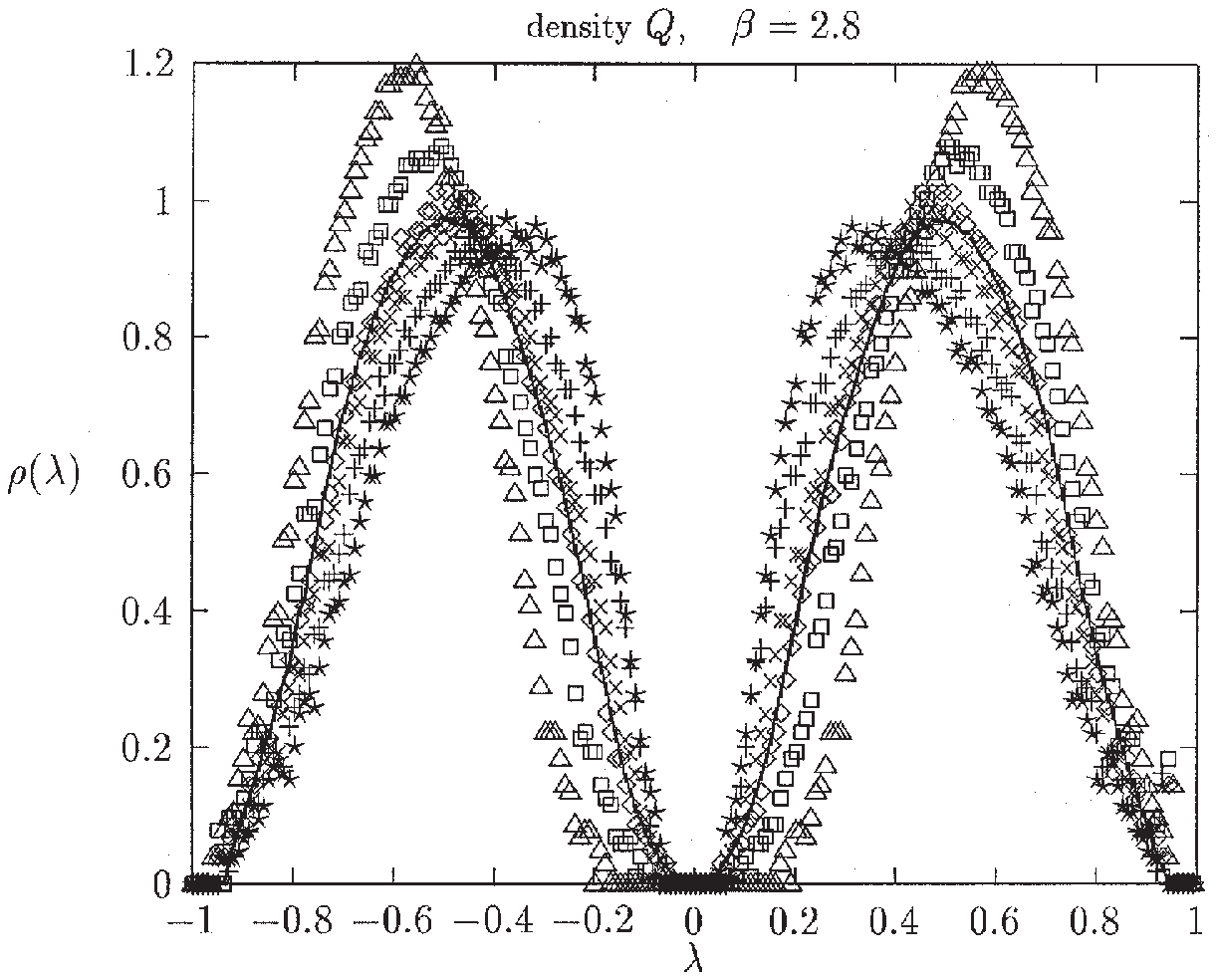,height=7.0cm} %,width=0.47\textwidth}
    \caption{Kalkreuter's results for $\beta=0.0$, $1.8$, and $2.8$,
      respectively. The eigenvalues are normalized to lie between $-1$
      and $1$. The symbols $\star$, $+$, $\times$, $\Diamond$, $\Box$,
      and $\triangle$ correspond to quenched data with $\kappa$ values
      of $0.25$, $0.20$, $1/6$, $0.15$, $0.125$, and $0.10$,
      respectively. The solid line shows unquenched data at
      $\kappa=0.15$ and $\beta=2.12$.}
    \label{beta18}
\end{figure}
One can clearly see the splitting of the spectral density
$\rho(\lambda)$ into two distinct symmetric parts at a critical value
of $\kappa$ between $0.2$ and $0.25$ as well as a slight dent on top
of the half-circle like densities. The structures come about as
follows.  The separation of the two half-circular structures is due to
the diagonal term $1/2\kappa$. The splitting of the two peaks for each
half-circle like structure results from the constant `$1$' in
\eqref{Dsl}. These dependences are in good agreement with the data.
The matrices $A$ and $B$ fluctuate according to the random matrix
constants $\Sigma_A$ and $\Sigma_B$ (see \eqref{Z}). If $\Sigma_A$ and
$\Sigma_B$ are large the structure gets washed out. If $\Sigma_A$ and
$\Sigma_B$ are different the half-circle like structures become
asymmetric.

Since the gauge fields in the action are completely replaced by random
variables the best agreement has to be expected for $\beta=0$. The
comparison shows that in this limit of extremely strong coupling RMT
is actually able to fit the global eigenvalue spectrum surprisingly
well. A detail which is missed is the slight difference in height of
the two maxima for, e.g., positive $\lambda$. Also the falloff for
large eigenvalues is steeper for the lattice results.

With decreasing coupling strength the agreement for the global
spectrum becomes worse, as is seen by comparing the left hand side of
Fig.~\ref{fig} to the lower two pictures of Fig.~\ref{beta18}, and one
reaches the usual situation, in which RMT can only describe the
microscopic fluctuations.

Let us conclude: We defined for the first time a random matrix model
which is suitable for the description of the eigenvalue spectrum of
the Dirac operator (multiplied by $\gamma_5$) for an SU(2) gauge
theory with Wilson fermions. A similar model should allow to study,
e.g., the distribution properties of the lowest eigenvalues for
SU(3)-Wilson fermions \cite{Jansen} and for the operator $(\Dsl+m)$
instead of $\gamma_5(\Dsl+m)$, which is a problem of great practical
importance.

\textbf{Acknowledgments.}
We thank T.~Wettig for very helpful discussions. This work has been
supported by BMBF.

\end{document}